\documentclass[aps,twocolumn,showpacs,amsmath,amssymb,superscriptaddress,final,10pt,prl]{revtex4}

\usepackage{graphicx}
\usepackage{color}
\usepackage{dcolumn} 
\usepackage{bm}      
\usepackage{nicefrac}
\usepackage{mathrsfs}

\newcommand{\BCPO}{Bi\-Cu$_2$\-PO$_6$}
\newcommand{\BCPOd}{Bi\-(Cu$_{1-x}$\-Zn$_x$)$_2$\-PO$_6$}
\begin{document}

\title{Field-induced quantum soliton lattice in a frustrated two-leg spin-\nicefrac{1}{2} ladder}

\author{F.\ Casola}
\email{fcasola@phys.ethz.ch}
\affiliation{Laboratorium f\"ur Festk\"orperphysik, ETH H\"onggerberg, CH-8093 Z\"urich, Switzerland}
\affiliation{Paul Scherrer Institut, CH-5232 Villigen PSI, Switzerland}%

\author{T.\ Shiroka}
\affiliation{Laboratorium f\"ur Festk\"orperphysik, ETH H\"onggerberg, CH-8093 Z\"urich, Switzerland}
\affiliation{Paul Scherrer Institut, CH-5232 Villigen PSI, Switzerland}%

\author{A.\ Feiguin}
\affiliation{Department of Physics, Northeastern University, Boston, Massachusetts 02115, USA}

\author{S.\ Wang}%
\affiliation{Laboratory for Developments and Methods, Paul Scherrer Institut, CH-5232 Villigen PSI, Switzerland}%
\affiliation{Laboratory for Quantum Magnetism, Ecole Polytechnique F\'ed\'erale de Lausanne, CH-1015 Lausanne, Switzerland}

\author{M. S. Grbi\'c}%
\affiliation{University of Zagreb, Faculty of Science, Physics Department, P.O. Box
331, HR-10002 Zagreb, Croatia}
\affiliation{Laboratoire National des Champs Magn\'etiques Intenses, LNCMI - CNRS (UPR3228), UJF, UPS and INSA,
BP 166, 38042 Grenoble Cedex 9, France}

\author{M.\ Horvati\'c}%
\affiliation{Laboratoire National des Champs Magn\'etiques Intenses, LNCMI - CNRS (UPR3228), UJF, UPS and INSA,
BP 166, 38042 Grenoble Cedex 9, France}

\author{S.\ Kr\"amer}%
\affiliation{Laboratoire National des Champs Magn\'etiques Intenses, LNCMI - CNRS (UPR3228), UJF, UPS and INSA,
BP 166, 38042 Grenoble Cedex 9, France}

\author{S.\ Mukhopadhyay}%
\affiliation{Laboratoire National des Champs Magn\'etiques Intenses, LNCMI - CNRS (UPR3228), UJF, UPS and INSA,
BP 166, 38042 Grenoble Cedex 9, France}

\author{C.\ Berthier}%
\affiliation{Laboratoire National des Champs Magn\'etiques Intenses, LNCMI - CNRS (UPR3228), UJF, UPS and INSA,
BP 166, 38042 Grenoble Cedex 9, France}

\author{H.-R.\ Ott}%
\affiliation{Laboratorium f\"ur Festk\"orperphysik, ETH H\"onggerberg, CH-8093 Z\"urich, Switzerland}
\affiliation{Paul Scherrer Institut, CH-5232 Villigen PSI, Switzerland}%

\author{H.\ M.\ R\o nnow}%
\affiliation{Laboratory for Quantum Magnetism, Ecole Polytechnique F\'ed\'erale de Lausanne, CH-1015 Lausanne, Switzerland}

\author{Ch.\ R\"uegg}%
\affiliation{Laboratory for Neutron Scattering, Paul Scherrer Institute, CH-5232 Villigen PSI, Switzerland}%
\affiliation{DPMC-MaNEP, University of Geneva, CH-1211 Geneva, Switzerland}%

\author{J.\ Mesot}
\affiliation{Laboratorium f\"ur Festk\"orperphysik, ETH H\"onggerberg, CH-8093 Z\"urich, Switzerland}
\affiliation{Paul Scherrer Institut, CH-5232 Villigen PSI, Switzerland}%

\date{\today}

\begin{abstract}
The field-induced quantum phase transitions (QPT) of the spin ladder material \BCPOd\ have been investigated via $^{31}$P nuclear magnetic resonance (NMR) on single-crystal samples with $x=0$ and $x=0.01$. Measurements at temperatures between 0.25 K and 20 K in magnetic fields up to 31 T served to establish the nature of the various phases. In \BCPO, an incommensurate (IC) magnetic order develops above a critical field $\mu_0 H_{c1} \simeq 21$ T; the field and temperature dependences of the NMR lines and the resulting model for the spin structure are discussed. Supported by results of Density-Matrix Renormalization Group (DMRG) calculations it is argued that the observed field-induced IC order involves the formation of a magnetic-soliton lattice. An additional QPT is predicted to occur at $H > H_{c1}$. For $x=0.01$, this IC order is found to be stable against site disorder, although with a renormalized critical field.
\end{abstract}



\maketitle

Quantum domain walls (QDW) are an important concept for understanding the magnetism of low-dimensional systems \cite{Villain75_Fad81_Coldea10}. 
In one dimension (1D), QDWs are due to the fractionalization of a spin-flip excitation into two spin-\nicefrac{1}{2} objects and represent kinks separating two energetically degenerate domains. Since their description involves non-linear differential equations \cite{Mikeska91}, they are usually termed $solitons$, in analogy to those known in condensed matter in relation with charge- and spin-density waves \cite{Brazovskii07,Blinc81}, lattice dislocations \cite{Bishop78} or superfluid $^3$He \cite{Maki76}. In 1D spin-\nicefrac{1}{2} antiferromagnetic (AFM) Heisenberg chains with strong geometrical frustration or spin-phonon interactions, QDWs are non-zero energy excitations that divide the chain into two regions with different dimerization patterns \cite{Sorensen98,Khomskii96,Chitra97}. 
Interchain interactions can select a dimerization pattern \cite{Khomskii96}, whose dimer bonds can be broken by applying a magnetic field exceeding a critical value $H_c$. A regular lattice of static QDWs then appears. 
The interest in non-linear phenomena and the related non-trivial excitations motivates the search for solitons induced solely by strong magnetic frustration,  a natural cause of dimer order \cite{Furukawa10,Bursill95}. 
Field-induced (FI) QDWs were characterized in detail via NMR and neutron diffraction in the prototype spin-Peierls system CuGeO$_3$ \cite{Mladen98,Henrik99}. The scarcity of model materials in which dimerization is triggered exclusively by magnetic frustration has so far limited the number of related experimental and, to some extent, theoretical studies. 

Based on results of experiments and model calculations presented below, we argue that the spin-\nicefrac{1}{2} zigzag ladder compound \BCPO, with geometric frustration caused by the next-nearest neighbor (NNN) interactions along the ladder legs, exhibits FI incommensurate (IC) magnetic order involving the formation of a QDW lattice. Upon replacing Cu by Zn the ordered phase persists, but is distinctly affected by moderate quenched disorder.

\begin{figure}[b]
\includegraphics[width=0.4\textwidth]{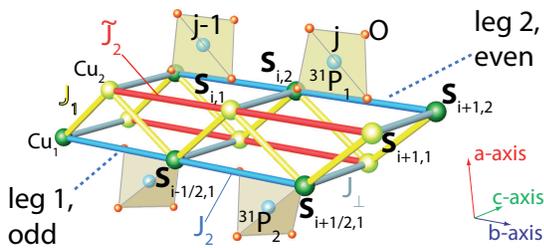}
\caption{\label{fig:structure}Structural unit of a spin ladder in \BCPO, including the positions of $^{31}$P nuclei. Only Phosphorus sites with $p=1,2$ are displayed. See text for details.}
\end{figure}

 {In \BCPO , the ladders formed by two coupled NNN frustrated chains run along the $b$-axis of the crystal structure \cite{Kote07,Shuang10}.} The inequivalence of the Cu$^{2+}$ sites allows the NNN exchange constants to alternate in magnitude \cite{Tsirlin10} ($J_2$ and $\tilde{J}_2$ in Fig.~\ref{fig:structure}). Residual interladder interactions are expected along the $c$-direction, while PO$_4$ tetrahedra act as non-magnetic spacers between different $bc$-planes. 
The strong spin frustration \cite{mentre09} and the unexpected observation of a complicated set of first- and second-order FI transitions above the closure of the spin gap $\Delta$ \cite{Tsirlin10,Kohama12} are reasons for the growing interest in this magnetic insulator.

\begin{figure}[t!]
\includegraphics[width=0.5\textwidth]{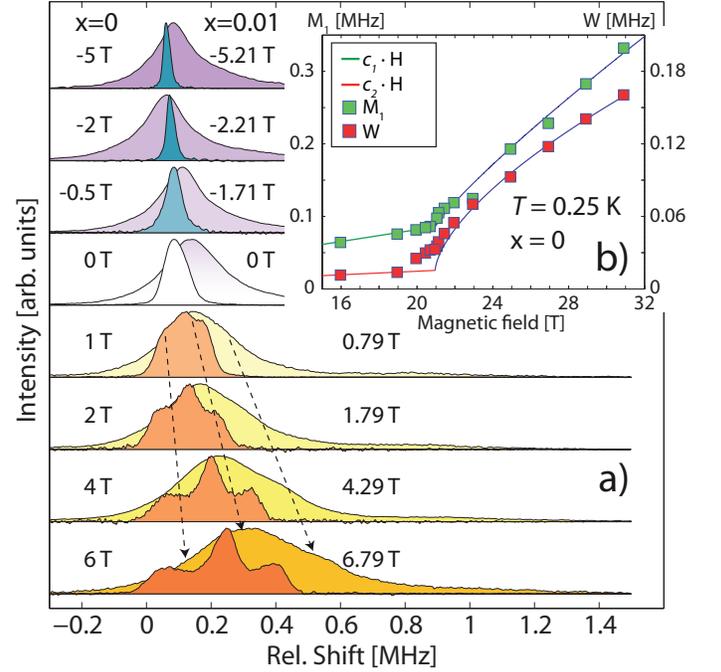}
\caption{\label{fig:Lines} (Color online) a) NMR lines $I(f)$ in the field-induced phase of \BCPOd . Lines are normalized to their peak maxima and their position is relative to the Larmor frequency. Data for both $x=0$ (foreground) and $x=0.01$ (background) at 
0.25 and 1.4 K, respectively, are displayed. Magnetic field values, reported to the left (right) of the line for the $x=0$ (0.01) case, are relative to the critical fields [$20.96(7)$ T and $24.21(9)$ T for $x=0$ and 0.01]. Arrows indicate the shift of the three blurred peaks for $x=0.01$. b) Field dependences of the first ($M_1$) and the square root of the second moment ($W$) as computed from the lines in (a) for the $x=0$ case. Blue curves are 
guides to the eye, while $c_1=4.1 \times 10 ^{-3}$ MHz/T and $c_2=0.73 \times 10 ^{-3}$ MHz/T.}
\end{figure}

In the first part of this Letter we present $^{31}$P NMR data obtained from $2\times 2 \times 2$ mm$^3$ single crystals \cite{Shuang10} of \BCPOd\ in magnetic fields up to 31 T, a range currently not accessible to, e.g., neutron diffraction experiments. The field was applied along the $b$-axis with $\pm$3$^{\circ}$ uncertainty and the covered temperature range was between 0.25 K and 20 K. 
The field dependence of the $^{31}$P NMR lines in \BCPOd\ for $x=0$ and $x=0.01$ is shown in Fig.~\ref{fig:Lines}a \footnote{The actual Zn content, measurable directly via NMR \cite{Casola10}, is found to be $x = 0.0081(5)$ for the sample investigated.}. We first consider the case of \BCPO. For $H<H_{c1}$ the system is disordered due to 1D quantum fluctuations and the regular local magnetization results in the same sharp resonance for all the $^{31}$P sites.
For $H>H_{c1}$ instead, a triple-peak NMR lineshape appears. It indicates a continuous distribution of local fields, suggesting magnetic order with an \textit{incommensurate} spin pattern \cite{Blinc81}. In general, the Cu$^{2+}$ electronic spin $\mathbf{S}_{i,l}$ contributes to the local-field experienced by the NMR nuclei via a priori unknown hyperfine interactions. In our case, a reliable evaluation of all these couplings was obtained by fitting the rotation patterns of the maxima of the main and satellite ${}^{31}$P NMR lines for $x=0.01$ \cite{CasolaNext,Casola10}. Quantitative comparisons of spin-structure models with experimental data are thus possible. The four ${}^{31}$P sites present in a unit cell are labelled by the index $p$. For $\mathbf{H} \parallel b$, the local field due to the $i$-th Cu${}^{2+}$ ion at the $j$-th ${}^{31}$P along the $b$ direction can, neglecting a minor dipolar contribution, be approximated by $h_p(j) \simeq \sum_{l,\lambda} A^{\lambda}_l(S^{\lambda}_{i+1,l}-S^{\lambda}_{i,l}) + \sum_{l} A^{b}_l(S^{b}_{i+1,l}+S^{b}_{i,l})$, where $\lambda = a,c$, the ladder leg is denoted by $l$ and the $i$ values are integer multiples of \nicefrac{1}{2} (see Fig.~\ref{fig:structure} for the notation). The hyperfine couplings $A^{\lambda}_l$ change sign according to symmetry when varying the index $p$. The longitudinal coupling $A^{b}_l \simeq 0.2$ T/$\mu_{\mathrm{B}}$ is dominant. Since $A^{a/c}_l \ll A^b_l$  \cite{CasolaNext}, the NMR spectra reflect mostly the longitudinal spin component $S^{b}_{i,l}$. The normalized NMR line\-shape $I(f)$ is dictated by the distribution of $\gamma_{\mathrm{P}} h_p(j)$, with $\gamma_{\mathrm{P}} H$ the $^{31}$P Larmor frequency. 
The onset of FI order at $H_{c1}$ is manifest through the strong increase of both the first moment $M_1$ and $W$, the square root of the second moment 
of $I(f)$ with field (see Fig.~\ref{fig:Lines}b) \footnote{By definition $ M_1= \protect\int_{-\infty}^{+\infty}  f I(f)\, \protect\mathrm{d}f = \gamma_{\mathrm{P}} N^{-1} \protect\sum_{p,j} h_p(j)$ and $W=\sqrt{M_2}=[\protect\int_{-\infty}^{+\infty}(f-M_1)^2I(f)\, \protect\mathrm{d}f]^{1/2} = [\gamma^2_{\mathrm{P}} N^{-1} \protect\sum_{p,j} h^2_p(j) - M^2_1]^{1/2}$, with $N$ the total number of $j$-sites}. 
 For $H<H_{c1}$ both $M_1$ and $W$ show a linear increase with $H$, suggesting that $S^{b}_{i,l}$ and $S^{\lambda}_{i,l} \neq 0$. In \BCPO , the $b$-axis is a principal axis for the $\mathbf{g}$-tensor and a possible Dzyaloshinskii-Moriya (DM) interaction 
associated with the $J_{\perp}$ coupling must be parallel to $b$ \cite{Tsirlin10}. If only these two terms are considered, the $U(1)$ symmetry \cite{ruegg08} is preserved for $\mathbf{H} \parallel b$, leading to $M_1 = W = 0$ for $H<H_{c1}$ \cite{Miyahara07}. 
The fact that this does not occur means that a DM interaction $\textbf{D}$ associated with the $J_1$, $J_2$ or $\tilde{J}_2$ exchange paths 
is the leading anisotropic term.

\begin{figure}[t]
\centering
\includegraphics[width=0.49\textwidth]{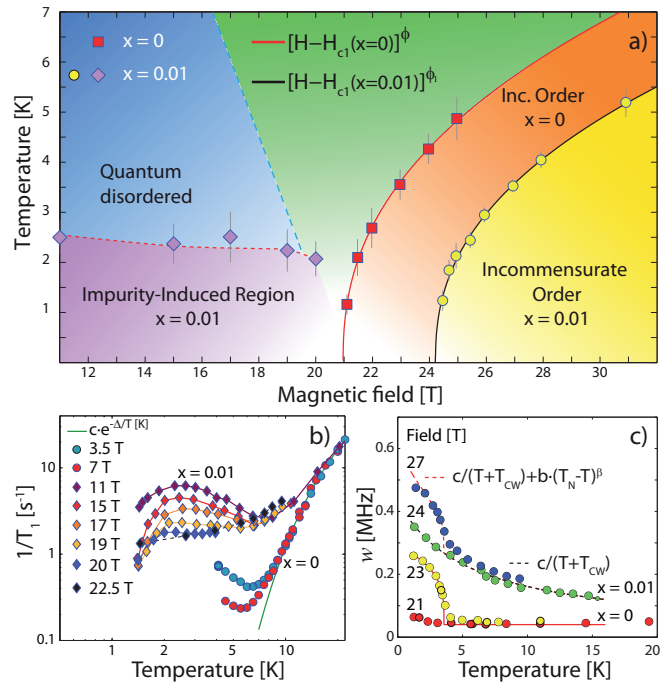}
\caption{\label{fig:PhaseBoundary} (Color online) a) Phase boundaries of \BCPO\ and \BCPOd\ for $H \parallel b$. Points are determined from $T^{-1}_1(T,H)$ relaxation data (see text for details), while $\phi = 0.42(5)$ and $\phi_1 = 0.41(3)$. b) $T$-dependence of $T_1^{-1}$ for $H<H_{c1}$, for $x=0$ (dots) and $x=0.01$ (diamonds). The solid green line is a fit to $c \exp(-\Delta/T)$, with $\Delta = 51.6(7)$ K. c) $T$-dependence of the NMR line widths $w(T,H)$ at $1/4$ of the line maxima for $x=0$ and $x=0.01$. The expression $c/(T+T_{\mathrm{CW}})$ represents a Curie-Weiss law, while the power-law exponent of the order parameter [fitted by fixing $T_{\mathrm{N}}$ from panel (a)] was found to be $\beta=0.31(6)$ for the $x = 0$ case.}
\end{figure}

The onset of magnetic order is also indicated by distinct maxima in the nuclear relaxation rate $T^{-1}_1(T,H)$ \cite{CasolaNext}, whose $(T,H)$-dependences are displayed as red squares in the phase diagram of Fig.~\ref{fig:PhaseBoundary}a.
The corresponding boundary was fitted using $T_{\mathrm{N}} \propto [H_c(T)-H_{c1}(0)]^{\phi}$, resulting in $\phi = 0.42(5)$; at $T=0$ K the onset of the FI order is at a critical field of $\mu_0 H_{c1}(0) =$ 20.96(7) T. The value of the exponent $\phi$  
is in good agreement with a recent mapping of the phase diagram using torque magnetometry at $T<1$ K \cite{Shuang2012}. Departures of $\phi$ from $\phi=2/d$, expected for $d$-dimensional Bose-Einstein condensation of triplet excitations in FI order of quantum AFMs, may be caused by both anisotropy and frustration \cite{ruegg08,Garlea09}. 

 The random suppression of spin degrees of freedom (known as \textit{site-disorder}) via the introduction of spin-zero Zn for Cu at a few-percent level has a significant impact on all the regions of the \BCPO\ phase diagram, namely, the quantum-disordered region [$k_{\mathrm{B}}T\leq \Delta(H)$], the IC phase [$H\geq H_{c1}(T)$], and the quantum-critical point [$H_{c1}(x,T=0)$]. As shown in Fig.~\ref{fig:PhaseBoundary}a, for $x=0.01$ the critical field increases by about 3.2 T to $\mu_0 H_{c1}(x=0.01,0) =$ 24.21(9) T. Contrary to the case of \textit{bond-disorder} \cite{Wulf11,Hong10,Yu12}, site-disorder allows for directly probing  finite-size effects.  For \BCPOd\ a larger spin gap is the result of the finite-size ladders between the Zn sites, hence implying a higher critical field $H_{c1}(0.01,0)$. FI magnetic order is absent in other disordered  quasi-1D spin system with incommensurate correlations \cite{Wulf11,Kiryukhin96}. 
Here, its existence can be demonstrated via NMR by monitoring the width $w(T,H)$ of the NMR lines at $1/4$ of the line maxima, which we choose as a measure of the order parameter. In Fig.~\ref{fig:PhaseBoundary}c we compare data obtained for $H-H_{c1}\simeq 0$ and 2 T, respectively. The field-induced contribution is well visible above the Curie-Weiss type broadening. A triple-peak line shape, albeit blurred due to disorder, is also observed for $x = 0.01$ (Fig.~\ref{fig:Lines}a). It indicates that the ladder regions between two Zn-sites tend to develop the same kind of IC order as for $x=0$. Supporting this conclusion is also the exponent $\phi$ which, in the measured temperature range and within experimental error, is found to be the same for both $x=0$ and  $x=0.01$.\\
Regarding the impurity-induced phase marked by diamonds in Fig.~\ref{fig:PhaseBoundary}a, we refer to $T^{-1}_1(T,H)$ shown in Fig.~\ref{fig:PhaseBoundary}b.
For $x=0$, an activated behavior reflects the gap in the excitation spectrum. Below $T = 10$ K there is a distinct upturn of $T_1^{-1}$, which is suppressed by the applied field. This can be related to residual impurity effects, as those occurring in gapped Haldane chains \cite{Fujiwara93}. 
For $x = 0.01$, with \textit{controlled} defects in the material, the same upturn is more pronounced. Here we observe maxima of these spin fluctuations occurring at a field-independent temperature, that disappear above 20~T. Most likely, the high-field IC-ordered phase is not directly adjacent to the low-field impurity-induced region, contrary to 
other Zn-doped cuprates possessing a spin gap \cite{Fujisawa06}. This opens the possibility to experimentally investigate the new plateau phases proposed for site disorder \cite{Yu10}. 


In what follows we use numerical tools to study the IC spin structure realized in \BCPO . We investigate, for $H \gtrsim H_{c1}$, the magnetization process of the ladder model with NNN exchange interactions proposed for \BCPO\ \cite{Tsirlin10} and show that it should not be seen as a simple increase of the number of individual bosonic-like spin-1 triplon excitations, but rather as the formation of quantum domain walls (QDWs) with effective spin \nicefrac{1}{2}. With our model, we can account for an additional QPT at $H > H_{c1}$ recently observed in experiments \cite{Kohama12}.

\begin{figure}[!t]
\centering
\includegraphics[width=0.5\textwidth]{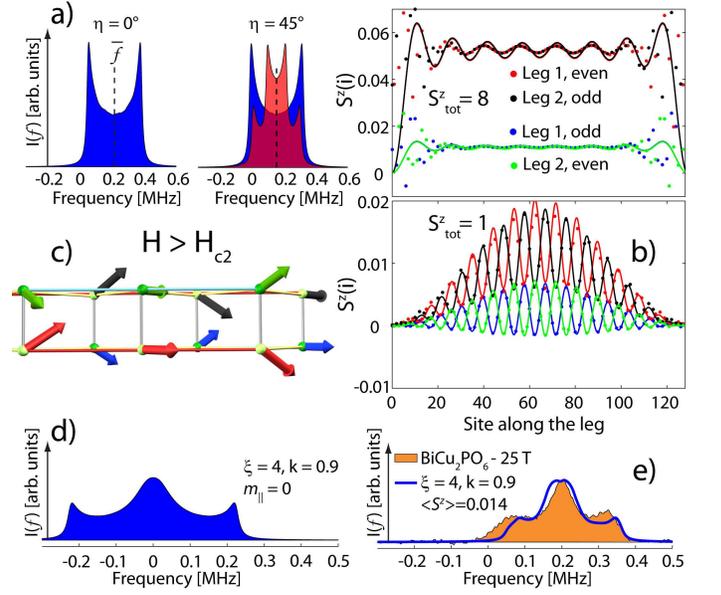} 
\caption{\label{fig:DMRG} (Color online) a) Simulated $I(f)$. Left: with $\eta,\epsilon = 0$ and $q_1 = 0.5 \pm \delta$ ($\delta \approx 0.04$ and $q$ is in reciprocal lattice units). Right: with $\epsilon = 0$, $\eta = 45^{\circ}$ and $q = q_1$ (blue) or $q = 2q_1$ (red). $m_{\perp}$ was taken such that the widths are comparable with the data in panel $e$. b) Points are $S^z(i)$ values from DMRG, while solid lines are fits to the free particle state $S_{\mathrm{free},l}^z(i_P)$ ($S^z_{tot} = 8$) and soliton $S_{\mathrm{sol},l}^z(i_P)$ state ($S^z_{tot} = 1$). c) Top view of the ladder in the C-phase predicted via DMRG calculations for $H>H_{c2}$ and $\mathbf{D}=0$. d) Expected NMR lineshape for a soliton lattice with $m_{l,\parallel}=0$ and $m_{l,\perp} \neq 0$ and $\eta,\epsilon=0$. e) Soliton lattice as in d), but with mean moment $\langle S^z \rangle \neq 0$ and $B_{l,p}=0.5$.}
\end{figure}

 From a \textit{classical} point of view, the natural state of frustrated 1D materials in a magnetic field is a canted spin helix \cite{Bursill95,Garlea08}. In a quantum-mechanical treatment,  a renormalization of both the pitch angle and the magnitude of the ordered moments is expected \cite{Enderle05}. Since in \BCPO\ the geometrical frustration acts only along the ladder, we considered a tilted cone arrangement of the magnetic moments with 
$\mathbf{S}_{i,l} = \mathbf{\bar{R}}_{\eta,\epsilon} (m_{l,\perp}\cos \theta_i, m_{l,\parallel}, m_{l,\perp}\sin \theta_i)$. The matrix $\mathbf{\bar{R}}_{\eta,\epsilon}$ rotates the spin structure away from the applied-field axis $b$ by a polar and an azimuthal angle ($\eta$, $\epsilon$), while $\theta_i = q i$, with $\mathbf{q} \parallel b$ as the propagation vector. For the ordered magnetic moments $m_{\perp}$, perpendicular to the cone axis, we used $m_{1,\perp} = -m_{2,\perp}$. The resulting NMR lines, simulated for several $\eta$, $\epsilon$, and $q$ values, using the known hyperfine parameters, exhibit 
profiles as those in Fig.~\ref{fig:DMRG}a, in clear disagreement with experimental data.   
The experimental resolution was taken as the width of the \BCPO\ line at 19 T; the calculated line shift is $\bar{f} = 2 \gamma_{\mathrm{P}}(A^b_1+A^b_2) m_{\parallel} \cos\eta$.

In order to retain quantum effects in the model, we performed a DMRG analysis of the FI phase transition in a system modeled by the Hamiltonian of two coupled frustrated $J_1$-$J_2$ chains, previously suggested for \BCPO\ in Ref.~\onlinecite{Tsirlin10} (with $J_1={J}_2$, $\tilde{J}_2={J}_2$/2, $J_\perp=3J_1/4$, $J_1/k_{\mathrm{B}}\simeq140$ K). We consider the low triplet-density limit, i.e., $H \gtrsim H_{c1}$, in systems with sizes up to $2 \times 128$. The complete study will be published elsewhere \cite{Adrian12}. 
The main results reported here turn out to be \textbf{qualitatively} different from those expected in the classical case. In a purely 1D system of length $L$, one can draw a convenient analogy between single triplet excitations and non-interacting hard-core bosons \cite{GiamarchiBook}. Therefore, with $n=S_{\mathrm{tot}}^z$ triplets in the finite system with open boundaries, the local magnetization is expected to be $S_{\mathrm{free},l}^z(i_P) = A_{l,P} \sum_{k=1}^{n}|\psi_k(i_P)|^2$ \cite{Fouet06}, where $\psi_k(i_P)$ is the particle-in-a-box wavefunction, as shown for the $S^z_{\mathrm{tot}} = 8$ case in the upper panel of Fig.~\ref{fig:DMRG}b. $P$ is the site parity, i.e., the $i_P$ values are even or odd multiples of \nicefrac{1}{2} and $A_{1,\mathrm{even}/\mathrm{odd}}=A_{2,\mathrm{odd}/\mathrm{even}}$. We call such a high-field phase commensurate (C) because $S_{\mathrm{free},l}^z(i_P)$ and therefore $m_{l,\parallel}$ is constant in the system if $L \rightarrow \infty$ and because, from numerical considerations \cite{Suppl12}, we expect in that regime a spin helix with $m_{l,\perp} \neq 0$ and a propagation vector close to the classical value $q_c =\pi^{-1} \arccos [-J_1/2(J_2+\tilde{J}_2)] = 0.608 \approx 2/3$ \cite{Adrian12}. The structure for  $q=2/3$ is sketched in Fig.~\ref{fig:DMRG}c. Approaching the critical field from above, $n$ decreases together with the overlap between particles, and the DMRG results predict a C-IC transition at an $L$-dependent value of $n$, corresponding to a second critical field $H = H_{c2}$. A continuous QPT is expected in case of weak interladder interactions \cite{Suppl12,Adrian12}.

In the IC phase, a triplet fractionalizes into two spin-\nicefrac{1}{2} objects. 
As shown in Ref.~\cite{Adrian12,Suppl12}, these QDWs break the short-range dimer order present along the ladder legs in the large $J_{\perp}$ limit \cite{Lavarelo11}, instead of the better known long-range dimer order studied at the Majumdar-Ghosh point of $J_1$-$J_2$ chains for instance \cite{Sorensen98}.
 By analogy with the $J_{\perp}=0$ case \cite{Sorensen98}, we fit the $S^z_{\mathrm{tot}} = 1$ data of Fig.~\ref{fig:DMRG}b with the expression $S_{\mathrm{sol},l}^z(i_P) = [1+B_{l,P}\sin(2\pi i q+\theta_{l,P})]S_{\mathrm{free},l}^z(i_P)$, where $q$ is incommensurate, $\theta_{l,\mathrm{even}}=\theta_{l,\mathrm{odd}}+\pi$ and $B_{1,\mathrm{even}/\mathrm{odd}}=B_{2,\mathrm{odd}/\mathrm{even}}$. In this model, the region of the phase diagram with FI order and $\mathbf{H} \parallel b$  investigated via NMR in the present work, corresponds to $H_{c1}\leq H \leq H_{c2}$. For the comparison with experimental data we need to describe this intermediate phase in the $L \rightarrow \infty$ limit. The latter is formed by commensurate regions between QDWs arranged in a regular pattern. For $H \gtrsim H_{c2}$, contrary to the IC region \cite{Adrian12,Suppl12}, the structure is a canted helix such as the one shown in Fig.~\ref{fig:DMRG}c. Hence we choose the equation for $\mathbf{S}_{i,l}$ to represent  the untilted cone ($\eta,\epsilon=0$) already used for the simulation in the left part of Fig.~\ref{fig:DMRG}a and use $\theta_i = \text{am}[i/(k \xi),k]$ for the phase modulating the transverse component (see also \cite{Suppl12}). In this notation, am is the Jacobi amplitude function, the solution of the sine-Gordon equation \cite{Abramowitz72} whose modulus is $k$. Plane waves are recovered for $k\rightarrow 0$ ($H \lesssim H_{c2}$), while the solitonic limit is reached at $k\rightarrow 1$ \cite{Blinc81} ($H \gtrsim H_{c1}$). For $m_{l,\parallel}=0$ and $m_{l,\perp} \neq 0$, the spin structure $\mathbf{S}_{i,l}$ exhibits zero magnetization, but the commensurate regions between so\-li\-ton phase-slips cause the predicted NMR profile to exhibit a triple peak (see Fig.~\ref{fig:DMRG}d). This line profile is qualitatively the same as those which were claimed to indicate soliton formation in charge-density-wave materials \cite{Koumoulis10,Blinc81}. The correlation length $\xi \approx 4$ is obtained via DMRG \cite{Suppl12}, while $k$ is close to one. To take into account the observed line shift of the NMR spectra, we have to consider also the longitudinal magnetization. We take $m_{l,\parallel}(i_P) = S_{\mathrm{sol},l}^z(i_P)$, replacing $S^z_{\mathrm{free},l}(i_P)$ with the periodic Jacobi function $A_{l,P}\text{dn}[i_P/(k \xi),k]$, which is a constant for $k\rightarrow 0$ \cite{Abramowitz72}. 
The resulting profile, as shown in Fig.~\ref{fig:DMRG}e, mimics satisfactorily the experimental data.
%

Based on the comparison between the observed and the calculated NMR lines, our interpretation of the IC phase applies to the field-induced order for small magnetic moments at $H_{c1}\leq H \leq H_{c2}$ and $\mathbf{H} \parallel b$. When the frustration-driven short-range dimer order is broken, a lattice of QDWs appears. Our model also predicts a \textbf{second}-order transition at $H_{c2}$ \cite{Adrian12} which, most likely, corresponds to the one observed in \BCPO\ at $H_2 \simeq$ 35 T using the same field orientation \cite{Kohama12}. Based on the preceding discussion, the IC-C character of the transition can be verified via future NMR experiments. For $\mathbf{H} \nparallel b$, the $\mathbf{g}$-tensor and the rung-DM anisotropy are more influential and therefore will have to be included to model the complexity of the overall phase diagram \cite{Kohama12}.

In summary, the field-induced order of the frustrated zigzag ladder \BCPOd\ has been investigated via $^{31}$P NMR with $H$ $||$ $b$. NMR data and DMRG calculations are combined to describe the magnetization process as the formation of a lattice of QDWs. For $x>0$ the persistence of the ordered phase, an enhanced critical field and enhanced low-temperature spin fluctuations in the low-field region are reported.

The authors would like to thank T.\ Giamarchi, P.\ Bouillot (Uni-Geneva), and M.\ Troyer (ETHZ) for numerous illuminating discussions. We thank  A.\ A.\ Tsirlin (MPI) for sharing his data prior to  publication. We acknowledge the EC support 
via the 7th framework programme ``Transnational Access'', Contract Nr.\ 228043\--Euro\-Mag\-NET II - Integrated Activities. This work was financially supported also by the Schweizerische Nationalfonds zur F\"{o}rderung der Wissenschaftlichen Forschung (SNF), the NCCR research pool MaNEP of SNF and the NSF Grant No. DMR-0955707.


\clearpage

\section*{SUPPLEMENTARY MATERIAL}
\subsection*{DMRG study of the soliton structure and of %
 com\-men\-su\-ra\-te-to-in\-com\-men\-su\-ra\-te transition in BiCu$_2$PO$_6$}

This is a short account of a DMRG numerical study of the ladder Hamiltonian introduced in Ref.\ \cite{Tsirlin10}. It serves to complement the main results presented in our work.
Here we focus on $(i)$ the nature of the C-IC transition, and $(ii)$ the structure of the correlations emerging as the spin system is partially polarized (at $H>H_{c1}$), with the ultimate goal of presenting a phenomenological picture that can be compared to experiments. All calculations (unless specified) were performed on two-leg spin ladders of up to 128 rungs, and using 600 DMRG states.

We start by recalling that when $\tilde{J_2}=0$, the geometry of the system shown in Fig.~1 of the main text, resembles that of two antiferromagnetically coupled sawtooth chains. Therefore, the physics of sawtooth chains, described in detail in Ref.\ \cite{Hao2011}, may well be closely linked to that of \BCPO. For example, 
the classical ground state expected in \BCPO\ for $\tilde{J}_2=J_2/2$ has an incommensurate spin order characterized by a wave vector $\mathbf{q}_c = (0.608,1)$ close to $2/3$, the value of coupled sawtooth chains \cite{Hao2011}.

At a quantum level, the sawtooth chain exhibits spontaneous dimerization \cite{Hao2011}. However, when the spin chains are coupled in the transverse direction to form a ladder, the rung interactions can dominate \cite{Lavarelo11}, and the ground state of the system is better described as being formed by strong \emph{rung singlets}. A strong rung-coupling regime has also been proposed for \BCPO\ \cite{Tsirlin10}.

\begin{figure}[t]
\includegraphics[width=0.45\textwidth,angle=0]{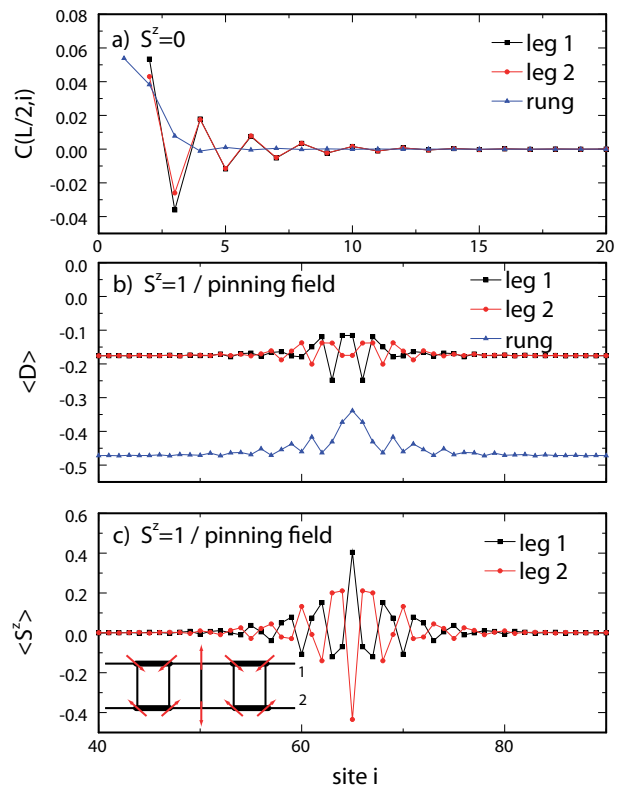}
\caption{a) Dimer-dimer correlations in the ground state of the \BCPO\ spin-ladder model described in the text, calculated relative to the center of the chain. The rung-rung correlations and the dimer-dimer correlations, both along the same leg and between opposite legs are shown. b) Dimerization pattern along the legs and rungs of a polarized ladder with a pinning field at the center of one leg. c) Spin texture surrounding the soliton structure. The cartoon in the inset illustrates the classical spin texture and the structure of the spin correlations (thick lines indicate stronger bonds).}
\label{fig:dimer}
\end{figure}

To study the system correlations 
we define the dimerization operators along the leg and rung directions as $D_{\mathrm{leg}}(\lambda,r)=\mathbf{S}_{\lambda,r} \cdot \mathbf{S}_{\lambda,r+1}$, and $D_\mathrm{rung}(r)=\mathbf{S}_{1,r} \cdot \mathbf{S}_{2,r}$, respectively, with $\lambda=1,2$ being the leg index. The dimer order parameter is then defined as $O_{\mathrm{dimer}}(r=i) = \langle D(i) \rangle-\langle D(i-1) \rangle$, with $i$ the site-index along the leg. From our calculations, both $\langle D_\mathrm{rung}(r) \rangle$ and $\langle D_\mathrm{leg}(r) \rangle$ are constant (except near the open boundaries), hence ruling out the presence of a spontaneous dimerization. In the unpolarized ($H < H_{c1}$) case, we find 
that $\langle D_\mathrm{rung} \rangle$ is much stronger than $\langle D_\mathrm{leg} \rangle$ (not shown), in agreement with Ref.\ \cite{Tsirlin10}. 
One can go one step further and study the dimer-dimer correlator $C(i,j) = \langle D(i) D(j) \rangle-\langle D(i) \rangle \langle D(j) \rangle$. Fig.~\ref{fig:dimer}a shows rapidly decaying rung-rung correlations (calculated from the chain center). This fast decay is consistent with a state of weakly correlated rung singlets. As also shown in Fig.\ \ref{fig:dimer}a, the dimer-dimer correlations along the leg direction indicate a cloud of sharply defined, but rapidly decaying staggered oscillations in a columnar pattern \cite{Lavarelo11}.

To break the rung singlets and create a nonzero magnetization, a critical magnetic field $H_{c1}$ is required. In this case, as shown in Fig.\ 4b of the main text, solitons are formed and an IC $S^z$ modulation is present. Intuitively, 
solitons can 
be visualized as spin-\nicefrac{1}{2} QDWs breaking the short-range leg dimer order present for $H<H_{c1}$ and visible in Fig.\ \ref{fig:dimer}a. Soliton pairs will then behave 
as triplet excitations with two ``up'' spins on the same rung. These pairs are weakly correlated along the leg direction and propagate coherently as hard-core bosons (see also Ref.\ \cite{Tsirlin10}). Solitons in \BCPO\ behave as weakly interacting extended objects, as shown in Figure 4b of the main text.

\begin{figure*}[t]
\includegraphics[width=0.5\textwidth,angle=-90]{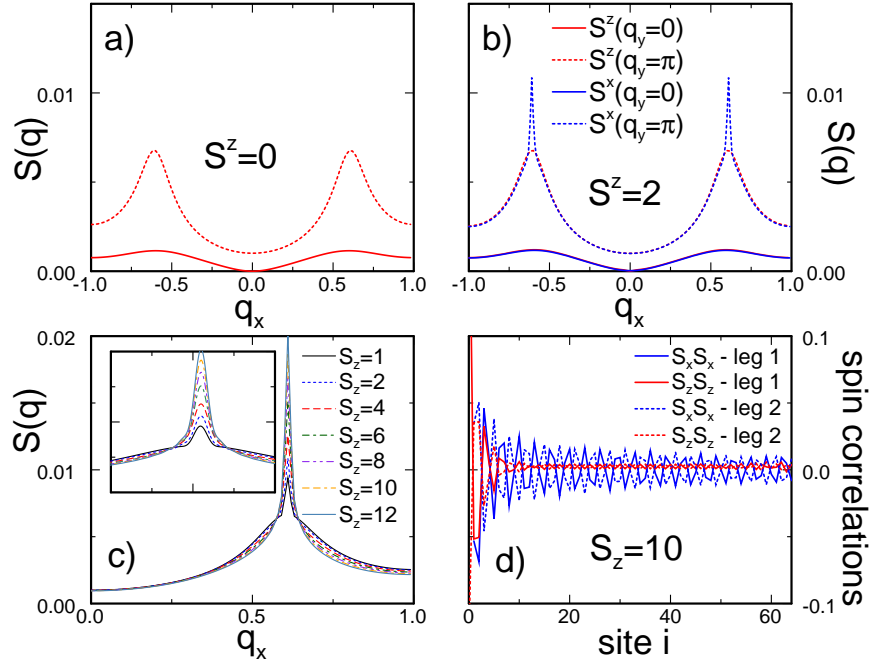}
\caption{Spin structure factor for a ladder with total spin $S^z=0$ (a) and $S^z=2$ (b). As magnetic field increases, a sharp peak develops in the transverse component, centered at $q=q_c$. The inset in (c) magnifies the region around $q_c$. Panel (d) shows the spatial structure of the correlations for $S^z=10$ as a function of the distance from the center of the chain. Notice that, in calculating the structure factor, the $\langle S^z(i) \rangle \langle S^z(j) \rangle$ contribution has been subtracted.}
\label{fig:sq}
\end{figure*}

{However, as we shall see below, the results from our DMRG simulations indicate that the detailed spatial 
structure of the domain walls is more complex}. A careful inspection shows the role of the incommensurate spin correlations with $q \approx 2/3$,  which are present as a quasi-classical spin texture superimposed on top of a dimerization pattern. To better characterize numerically the spatial profile of solitons, we added a pinning field at the center of the chain. The pinning field localizes a spin pointing in the ``up'' direction. Correspondingly, the spin on the same rung but 
opposite leg spontaneously points in the ``down'' direction. The surrounding background of spins arranges itself in
a classical structure with $\mathbf{q}=\mathbf{q}_c$. The excess spin ``up'' in the neighboring chain is not localized on the same rung, but it spreads along the second chain over the width of the soliton, preferentially on the two sites to the left and right of the center, at an angle of $\sim 120^\circ$ obtained from the value of $q_c$ (see inset of Fig. \ref{fig:dimer}c). 

The soliton structure emerging from 
our calculations is unusual, and differs from those 
present in systems with a commensurate spin order. Moreover, due to the different couplings ($\tilde{J}_2 \neq J_2$), the profile of the soliton depends on whether it is situated on even or on odd sites of the leg, similarly to what happens in the sawtooth chains. Notice that the reference position in our calculations corresponds to an odd site on one of the ladder legs. Results for even sites, although
quantitatively different, are qualitatively equivalent and hence are not shown here.

\begin{figure}[t]
\includegraphics[height=0.45\textwidth,angle=-90]{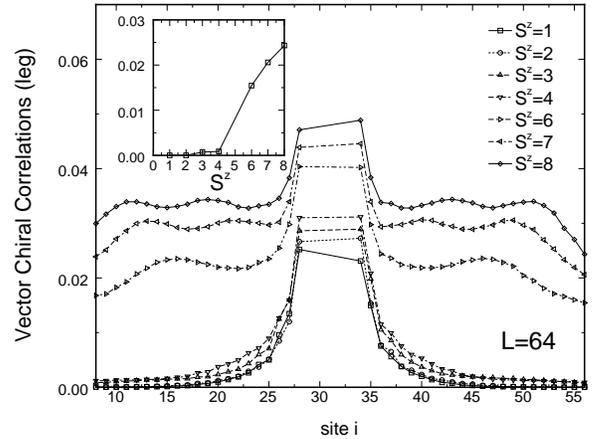}
\caption{Vector chiral correlations along a leg of the ladder, measured relative to the center, for a system of length $L=64$. The inset shows the magnitude of the correlations $10$ sites before the end of the chain, for different magnetizations.
}
\label{fig:CHIRAL}
\end{figure}

In Fig. \ref{fig:sq} we characterize the spin-spin correlations accompanying the FI phases. In the $S^z=0$ case, the system is quantum disordered with a Lorentzian-like spin structure factor $S(q)$ in reciprocal space (Fig.\ \ref{fig:sq}a). Once the critical field $H_{c1}$ is reached, $S^z>0$ and the 1D system starts to develop quasi-long range order. The latter is manifest in the calculations with a sharp cusp in the spin structure factor, centered at $q=q_c$ (see Fig.\ \ref{fig:sq}b and \ref{fig:sq}c). We also observe in Fig.\ \ref{fig:sq}d that the correlation functions decay more rapidly for the spin components longitudinal to the applied field than for those perpendicular to it. The latter is a common situation in the gapless phase of two-leg ladders \cite{Giamarchi99} and implies that in the presence of interladder interactions $J'$, magnetic order will first involve the perpendicular component \cite{Hikihara10}. It has been shown \cite{Lauchli08} that long-range vector chiral order (breaking a discrete parity symmetry $Z_2$) can develop even in 1D. A possible instability towards helical order in case of $J' \neq 0$ can be identified by numerically computing the chiral correlations $\langle \kappa(i) \kappa(j) \rangle$ with the chiral operator $\kappa_i =  (\mathbf{S}_{i} \times \mathbf{S}_{i+1})^z$. In Fig. \ref{fig:CHIRAL} we see that vector chiral correlations are \textbf{short-ranged} in the region $H_{c1}<H<H_{c2}$, whereas a more classical cone is the ordered state realized in the 3D case for $H>H_{c2}$.

For $H_{c1}<H<H_{c2}$, an incommensurate quasi-classical spin order develops around each QDW as the density of solitons increases. 
As in the case of CuGeO$_3$ \cite{Zang97}, the interactions among solitons are supposed to affect mainly the soliton-soliton distance (and, in turn, $m_{\parallel}$) but not the correlation length $\xi$. Therefore, at a critical density $m^c_{\parallel}$ corresponding to the inverse soliton width, namely $m^c_{\parallel} \approx 1/2\xi = 1/8$, where $\xi \approx 4$ is the correlations length, the solitons start to overlap and, due to the soliton-soliton repulsion, become more localized. In our system, we find that this corresponds to the development of long range chiral correlations at $H_{c2}$. 

 Due to simple energy considerations, the longitudinal magnetization parallel to the field develops 
predominantly on one of the sublattices (even sites on leg 1 and odd sites on leg 2 --- see Fig.\ 4b in the main text). The longitudinal magnetization is visible in the spin-spin correlations as a small commensurate feature with $q=(1,1)$. It is not seen, though, in the structure factor, since the $\langle S^z(i) \rangle \langle S^z(j) \rangle$ contribution has been subtracted. Notice that this term would also have lead to a peak proportional to $S^z_{tot}$ at $q=0$.

To bear out 
the features associated with the soliton structure, we have ignored the effects of DM interactions in the above discussion. Adding them would add another degree of complexity to the problem \cite{Hao2011}.

\end{document}